\documentclass[]{aa}
\usepackage{amsmath}
\usepackage[dvips]{graphics}
\usepackage {epsfig}
\title{Perturbative reconstruction of a gravitational lens: when mass does not follow light}
\author{C. Alard 
        \inst *}

\institute{Institut d'Astrophysique de Paris, 98bis boulevard Arago, 75014
           Paris \\
           \email{alard@iap.fr}}
\date{}
\begin{document}
\abstract {} {The structure and potential of a complex gravitational
  lens is reconstructed using the perturbative method
  presented in Alard 2007, MNRAS, 382L, 58; Alard 2008, MNRAS, 388, 375.
This lens is composed of 6 galaxies belonging to a small group.}
{The lens inversion is reduced to the problem of reconstructing
 non-degenerate quantities:
the 2 fields of the perturbative theory of strong gravitational lenses. 
Since in the
perturbative theory the circular source solution is analytical, the
general properties of the perturbative solution can be inferred directly from the data.
As a consequence, the reconstruction of the perturbative fields is not
affected by degeneracy, and
finding the best solution is only a matter of numerical refinement.} 
{The local shape of the
potential and density of the lens are
inferred from the perturbative solution, revealing the existence
of an independent dark component that does not follow light.}
{The most likely explanation is that the particular shape of the dark
halo is due to the merging of cold dark matter halos. This is a
new result illustrating the structure of dark halos at the scale of
galaxies.}
\maketitle
\section{Introduction.}
In 1986 Lynds and Petrosian reported the discovery of giant arcs in Abell 370 and 2242-02, with
a length of 100 Kpc and luminosities comparable to giant elliptical
galaxies, but with bluer colors.
Bohdan Paczy\'nski (1987) interpreted these features, he suggested that the arcs may be the result of the
clusters acting as gravitational lenses. The Paczy\'nski interpretation was later confirmed
by Grossman \& Narayan (1988), and initiated a series of very fruitful
developments on
gravitational lenses. The lens inversion problem was tackled 
 by Kochanek { et al.}
(1989) using the ring cycle method. Other
methods have developed to solve the lens inversion problem.
Such as maximum entropy methods by Wallington { et al.} (1996),
methods based on Fourier series expansion (Trotter { et al.} 2000),
LensCLEAN (Wucknitz 2004), semi linear inversion methods (Warren \& Dye 2003), Bayesian
methods  (Brewer \& Lewis 2006), and potential reconstruction method
(Suyu { et al.} 2009).
The lens inversion methods may be classified in 3 types: (i) model dependent reconstructions,
(ii) potential reconstruction on a grid, (iii) expansion of the
potential in Fourier series.
All methods have to consider the degeneracy of modeling, for instance
Wayth { et al.} (2005) found that
6 different models were consistent with the observations for the lens ER 0047-2808. The problem of
 degeneracy in lens inversion is widespread. The degeneracy issue is related to the fact that the constraints on the
 potential are local. The value of the potential is constrained only in areas where images of the source are
formed. Additional constraints comes from the fact that no images are formed in
dark areas (Diego { et al.} 2005). Since there are no
constraints on the potential in other areas, without additional assumptions, the problem is fully degenerate,
 and an infinite number of models are possible. One possibility is to explore the family of models that
is consistent with the observations (Saha \& Read, 2009) and to try to find the common features between the
models (non-degenerate quantities). However, in such an approach it is difficult to
explore the full range of possible models. 
Another possibility to reduce the degeneracy is to use a local representation of the potential, like
the methods (ii) and (iii). 
However, the grid methods or Fourier methods require many free parameters, and
it is not clear whether the degeneracy problem has been improved. To solve the degeneracy problem it is essential
to reduce the number of parameters, and to describe the potential in terms of fundamental, non-degenerate quantities.
Since giant arcs resemble a part of a circle, it is natural to consider that arcs corresponds to a small break
of the circular symmetry. In such an hypothesis, the fundamental properties of the arcs should be related to the
quantities breaking the circular symmetry. These ideas were first explored by Blandford \& Kovner (1988) who 
demonstrated that for a small impact parameter (with respect to the size of the critical circle) and small deviations
from circularity, the equation for caustics depends only on the orthoradial component of the deflection field
at the critical circle. Blandford \& Kovner (1988) also presented a geometrical method to reconstruct
the images of the lens. The same concepts were investigated later by Alard (2007) who derived an analytical
equation to describe the formation of images by the lens. The main advantage of the analytical description
is that it directly relates the morphological properties of the arcs to the potential expansion. This relation
 between the potential and the observational features suggests that the description of
the potential is non-degenerate, and also simplifies the lens inversion problem. The reduction of the
 degeneracy problem in this approach is well illustrated by the fact that an infinite 
number of models corresponds to a given perturbative representation. 
Lens inversion in this perturbative framework was already tested in Alard (2008) using numerical simulations.
It was found that the perturbative approach allows an accurate re-construction of the potential for lenses with
substructures. 
The concepts developed in this approach are very different
from some recent perturbative numerical schemes
(see for instance Vegetti \& Koopmans 2009). In this type of work
there is no attempt at reformulating the lens-inversion problem
in terms of fundamental quantities; the main goal is to derive
an efficient numerical scheme to solve the ordinary lens equation.
Peirani {et al.} (2008) showed that the perturbative 
approximation is accurate for realistic lens models. The numerical
accuracy is of about 1\% in units of the critical
radius. The lens model of Peirani {et al.} (2008) were extracted
from cold dark matter simulations (Horizon project), and some
of them have a lot of structure (merging). This paper is organized as follows: the first part 
will present the perturbative method, and its basic properties. In an effort to make this
work self-contained, a synthesis of the other useful results on the pertubative method is presented
in the appendices. This section also has the advantage of presenting all the equations essential
to lens inversion.
The second part of the paper will present the
application of the perturbative method to the inversion of an interesting 
case of gravitational lens. This paper also will be an occasion to extend   
the technique of perturbative inversion and to relate the perturbative
fields to the geometry of the lens.
\section{The perturbative approach in gravitational lensing.}
 In the perturbative approach (Blandford \& Kovner 1988 and Alard 2007), gravitational arcs
 are interpreted as a perturbation of the perfect ring situation.
  A perfect ring is formed when a point source is situated at the center of a circularly symmetric
 lens. In the perfect ring situation, the central 
  point has an infinite number of images located on the critical circle. 
There are 2 different kinds of perturbations that breaks the symmetry of the system: the
 non-circularity of the lens, and the source impact parameter.
When such non-circular perturbations are introduced, the symmetry of the system is lost and
  the ring is replaced with a series of individual images. However, as
 illustrated in Alard (2007) whatever the angular position $\theta$ of these
perturbed images, there is always an unperturbed image of the central
 point on the critical circle at the same angular position. 
As a consequence there is only a radial displacement $dr$ between the perturbed and un-perturbed images.
The un-perturbed situation is represented by a circular lens with potential $\phi_0(r)$, and
a point source with null impact parameter. In the perturbed situation, the source has
an impact parameter ${\bf r_S}$, and the lens is perturbed by the non-circular potential 
$\psi(r,\theta)$.
Both perturbations ${\bf r_S}$ and $\psi$ are assumed to be of the same order $\epsilon$:
  \begin{equation} \label{perts}
  \begin{cases}{\bf r_S} &= \epsilon \ {\bf r_s}  \\
   \phi &= \phi_0 + \epsilon \ \psi
   \end{cases}
  \end{equation}
 { Note that for convenience the unit of the coordinate is the critical radius. 
Thus by definition in this coordinate system
 the critical radius is situated at $r=1$.} As a consequence, in the
 continuation of this work all distances and their associated quantities (errors,..) will be
 expressed in units of the critical radius. 
 In response to this perturbation, the radial position of the image will be shifted by an amount $dr$
 with respect to the unperturbed point on the critical circle.
 The new radial position of the image is:
 \begin{equation}
  r = 1 + \epsilon \ dr 
 \label{r_def}
 \end{equation}
  The response $dr$ to the perturbation can be evaluated by solving the lens equation in the
  perturbative regime. The lens equation reads:
  \begin{equation}
   {\bf r_s} = \left(r
   -\frac{\partial \phi}{\partial r} \right) {\bf u_r} - 
   \left( \frac{1}{r} \frac{\partial \phi}{\partial \theta} \right) {\bf u_\theta}
  \label{lens_eq}
  \end{equation}
  The last step to solve the lens equation is to expand the potential is series of $\epsilon$:
  \begin{equation}
    \phi = \phi_0 + \epsilon \psi = \sum_{n=0}^{\infty} \left [ C_n +
    \epsilon \ f_n(\theta) \right] \ (r-1)^n \\
    \label{phi_pert}
   \end{equation}
with:
  \begin{equation}
   \begin{cases}
    C_n = \frac{1}{n!} \left[ \frac{d^n \phi_0}{ d r^n} \right]_{(r=1)} \\
    f_n(\theta) =  \frac{1}{n!} \left[ \frac{\partial^n \psi}{ \partial r^n} \right]_{(r=1)} \\
   \end{cases}
   \label{fc_defs}
   \end{equation}
  By inserting Eqs (~\ref{perts}), (~\ref{r_def}), and (~\ref{phi_pert}) in the lens equation (Eq. ~\ref{lens_eq})
  , it is straightforward to obtain an equation relating the response $dr$ to the perturbation $r_S$ and $\psi$:
 \begin{equation}
  {\bf r_s} = \left( \kappa_2 \ dr - f_1 \right) \ {\bf u_r} - 
 \frac{d f_0}{d \theta} \ {\bf u_{\theta}} 
 \label{final_pert_eq}
 \end{equation}
 and $\kappa_2=1-\left[ \frac{d^2 \phi_0}{ d r^2} \right]_{(r=1)}$ \\\\
 This equation corresponds to Eq. (8) in Alard (2007).

%%%%%%%%%%%%%%%%%%%%%%%%%%%%%%%%%%%%%%%%%%%%%%%%%%%%%%%%%%%%%%%%%%%%%%%%%%%%%%%

{ Considering that the source has a mean impact parameter ${\bf
    r_0}$, the position in the source plane may be re-written:
    ${\bf r_S=\tilde {r_S}+r_0}$. Assuming Cartesian coordinates $(x_0,y_0)$
    for the vector ${\bf r_0}$, Eq. ~\ref{final_pert_eq} reads:
\begin{equation}
 {\bf \tilde  r_s} = \left( \kappa_2 \ dr - \tilde f_1 \right) \ {\bf u_r} - 
 \frac{d \tilde f_0}{d \theta} \ {\bf u_{\theta}} 
 \label{eq_tilde}
\end{equation}
For a circular source with radius $R_0$, the perturbative response $dr$ takes the simple following form (Alard 2007, Eq. 12):

 \begin{equation} 
 dr = \frac{1}{\kappa_2} \left[ \tilde f_1  \pm \sqrt{R_0^2-\left( \frac{d \tilde f_0}{d \theta} \right)^2} \right]
 \label{dr_eq}
 \end{equation}
 }

Eq. (~\ref{final_pert_eq}) depends on $\kappa_2$. However
this variable can be eliminated from Eq. (~\ref{final_pert_eq}) by re-normalizing the fields:
$f_n=\frac{f_n}{\kappa_2}$, and the source plane coordinates, ${\bf
  r_s}=\frac{{\bf r_s}}{\kappa_2}$ (mass sheet degeneracy). These
re-normalized variables will be adopted in the rest of this
work. The re-normalization is equivalent to solving Eqs (~\ref{final_pert_eq}) and
(~\ref{dr_eq}) for $\kappa_2=1$. The variable
$\kappa_2$ will re-appear when the re-normalized quantities are
converted to the original quantities.
%
%%%%%%%%%%%%%%%%%%%%%%%%%%%%%%%%%%%%%%%%%%%%%%%%%%%%%%%%%%%%%%%%%%%%%%%%%%%%%%%%%%%%%%%%%%%%%%%%%%%%%%%%
%
\section{Pre-processing of the HST data.}
\begin{figure}[htb]
\centering{\epsfig{figure=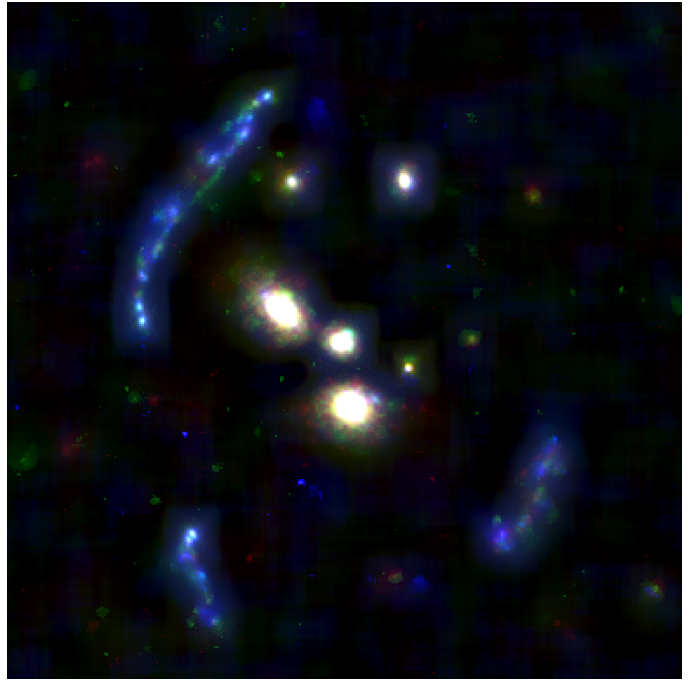,width=9cm}}
\caption{Color image for SL2SJ02140-0532 reconstructed from 3 noise
  filtered HST images in 3 bands.}
\label{plot_1}
\end{figure}

The gravitational lens SL2SJ02140-0532 (Cabanac { et al.} 2007,
SL2S public domain) was observed by HST in 3 
spectral domains, F475W, F606W, F814W, with an exposure time of 400
seconds. The main difficulty in the pre-processing of the images
is to remove the large number of cosmic present in the images. Cosmic
are identified using a wavelet approach based on an estimation of the local scale
in the image. Once a cosmic has been identified at a given position,
the local area around this position is investigated to remove any other
pixels connected to the cosmic. This procedure is based
on routines from the ISIS package (Alard 2000).
Note that since the number of cosmic is large in the images, the
local density of cosmic sometime become too large in some areas.
 When
too many nearby cosmic are detected within a small area, the area
is flagged, and will be considered cautiously, or even rejected.  
{ Once the 3 images are cosmic cleaned, a re-centering to a common grid is performed 
using the position of the bright objects. The last step in the processing of the images
is to remove the background.
%%%%%%%%%%%%%% Ici
 There is some diffuse light coming from the galaxy halos at
the level of the arcs, and the proper removal of this diffuse light is especially important.
The diffuse background is removed in several steps. The first step is to build a general
background model. To build this model, the image is divided
in square areas of 32 pixel, the local background is estimated within each area using
a $\kappa\sigma$ cleaning method. The  background value in each local area is then interpolated
to compute a general background value at each position in the image. 
Once this background model has been subtracted
from the image, the residuals around the arcs are investigated in
order to control any systematic errors.
A tiny systematic background residual is detected; this residual depends on the distance to the 3 main
galaxies center of gravity. This background residual is related to an imperfect subtraction of a 
diffuse halo centered around the 3 main galaxies. An elliptical component centered
at the center of gravity of the 3 galaxies is fitted to this residual. 
The best elliptical parameters are estimated by minimizing the
deviations to elliptical symmetry. The mean radial profile of the residual is estimated,
 and finally the relevant elliptical component is subtracted. Note that this last background
 correction is small, it is only a few percent of the initial background subtraction. 
}
Thus, the 3 cosmic cleaned, re-centered and
 background subtracted images are stacked,
to produce the final reference image of the arc system.  A color illustration
of the arc system is provided in Fig. ~\ref{plot_1}.
\section{Lens inversion.}
{ This work presents the first successful model of the images formed by the 
lens SL2SJ02140-0532. This system was investigated by Limousin {
  et al.} (2009),
who tried to fit this system using an elliptical lens model. They found that it was not possible
to reproduce this system using such simple elliptical models.
Limousin {et al.} (2009) were
also confused by the color of the last image (right side in Fig. ~\ref{plot_1}) which is slightly 
different to the color of the other images. They concluded that the identification of the images
formed by this lens was uncertain, and as a consequence renounced modelling of this system. We will
see in the continuation of this work that this lens requires models much more complex than elliptical,
and that such models can be built naturally using the perturbative method presented in this paper.
The development of this perturbative model also will naturally solve the problem of the different
color of the last image. We will see that the first part of this image has the same
color as the other images, while the remaining part of the image has a different color. This is a
direct prediction of the perturbative model, which offers an elegant solution to the problems encountered
by Limousin {et al.} (2009). We now turn to the analysis of this lens using the perturbative method.} 
The lens SL2SJ021408-053532 forms a system of 4 images of the source. The first 2 images
 are situated inside the large arc in the
upper left (they appear as 2 identical mirror images). The 2
 other image, 
are at the bottom and the right side of the frame (see Fig ~\ref{plot_1}).
In the perturbative approach the problem of reconstructing this lens is
equivalent to the reconstruction of the perturbative fields, and the next sections
will concentrate on this issue.
{ Obviously, the re-construction problem is reduced to the reconstruction of the 
first order perturbative field only if the first order perturbative
theory applies. The requirement for this approximation
to work is that the gradient of the potential can be linearized in the vicinity
of the critical circle (Alard 2007). For single dark matter halos numerical experiments 
shows that this local
constraint on the gradient is usually met with good accuracy (Alard 2007, Peirani {et al.} 2008).
It is only in the case where a local minima of the potential is present close to the critical
circle that this approximation may not work. But such situations are very rare because they require
that for instance a sub-structure falls very close to the critical line. However Alard (2008)
showed that the general treatment of substructure in this approximation is correct even when the substructure
is quite close to the critical circle (up to one tenth of the critical circle). Additionally, when an unlikely
 situation happens, such as a substructure lying on the critical line, the consequences are easily observable:
a few additional images are formed near the perturbator. Thus such situation is easy to recognize, and is not
apparent in the present data. In any case the best test of the first order approximation is to perform a successful
reconstruction of the data, which means that in practise second order terms can be neglected. This will
be the approach favored in this paper.
}
Before the field re-construction problem can be tackled, a necessary step
is to evaluate the size of the critical circle, which will become by convention
the coordinate system unit.
\subsection{Estimation of the critical circle.}
The critical circle is estimated by fitting a circle to the mean
position of the images. The center and radius of the circle are
adjusted. The non-linear adjustment procedure starts
from a circle centered on the small galaxy at the center of the image.
The initial estimate for the circle radius is the mean distance of the
images to the center. A non linear optimization shows that the best
center is close to the initial guess and that the optimal radius is
close to 152 pixels; with a pixel size of $0.049^{\prime \prime}$ the critical circle
radius $R_C$ is $7.44^{\prime \prime}$. This value should be compared to the radius
of $7.31^{\prime \prime}$ found by Limousin {et al.} for the same lens. Note that
the final result will not depend upon the particular choice of a given
critical circle. Taking another circle close to this one would change
a little the estimation of the perturbative fields, but the total
background plus perturbation would remain the same.
\subsection{General properties of the solution.}
\begin{figure}[htb]
\centering{\epsfig{figure=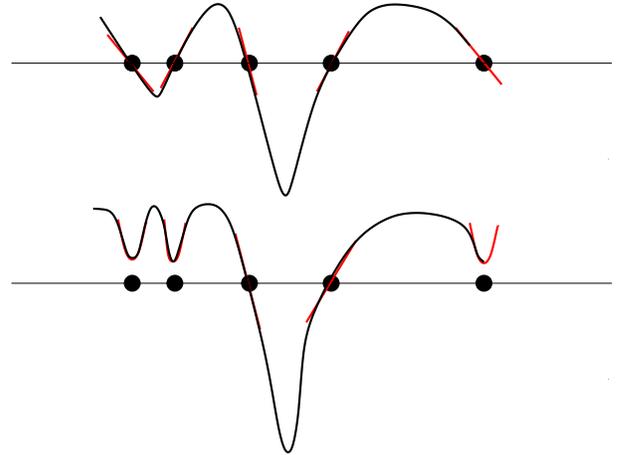,width=8cm}}
\caption{Two example of possible shapes for the field $\frac{d \tilde f_0}{d
    \theta}$. The black dots represents the mean image position. The
    red curve represents the local behavior of the field near the image, while the
    black curve is a smooth extrapolation. The
    upper curve present a solution where the field has local linear behavior
    near the images. In the lower curve a higher order behavior is
    presented. In dark areas the field must be larger that the source
    radius to avoid forming images (see Eq. ~\ref{dr_eq}). 
    Note also that the field  $\frac{d \tilde f_0}{d \theta}$ must also have zero sum.
}
\label{plot_topo}
\end{figure}

The general properties of the solution can be derived from the
properties of the circular source solution. To estimate
the circular source solution, the outer contour of the source will be
 approximated with a circle. This approximation of the source may not
 be very accurate; the typical error on the field estimation will be
 of the order of the source outer contour deviations from circularity.
However, this approximation is sufficient to obtain a first
estimate of the solution and to derive its  general
 properties. For circular sources
the perturbative fields are directly related to the data (see Appendices ~\ref{appendice_1},
and Eq. ~\ref{dr_eq}). This direct
 relation between the fields and the data is possible only in areas of the lens plane
where images of the source are present. In dark areas, the field
will have to be reconstructed by interpolating the values in bright
areas. The field $\tilde f_1$ will be
approximated with the mean radial position of the images (with respect
to the critical circle). The reconstruction of $\tilde f_1$ is simple and not
ambiguous, but the reconstruction of $\frac{d \tilde f_0}{d \theta}$ requires
more work. On the other hand $\frac{d \tilde f_0}{d \theta}$ is better
constrained since no images should
be formed in dark areas ($|\frac{d \tilde f_0}{d \theta}|>R_0$). 
Furthermore, the field $\frac{d \tilde f_0}{d \theta}$ is of
particular interest since it is directly related to the potential
iso-contour (Eq. (~\ref{iso_eq})).
The derivation of the general properties of
$\frac{d \tilde f_0}{d \theta}$ require some specific guess of the
local behavior of this field in the region of image formation. Images
form in minima of $|\frac{d \tilde f_0}{d \theta}|$ and the example
 presented in Appendix ~\ref{appendice_1}
indicates that the
 behavior of  $\frac{d \tilde f_0}{d \theta}$  near the minima is of 2 types.
 For small images  $\frac{d \tilde f_0}{d \theta}$ has a linear
 behavior, while
 for larger images (caustics) the field behaves like a higher order
 polynomial. In the case of this particular lens system, we have 4
 images of comparable size. The typical image size is quite small
 and it is reasonable to assume that the field behavior in the image
 is linear. Of course other models could be investigated, with for
 instance order 2 behavior of the field, but obviously these models
 are much more complex and would require higher order Fourier
 expansions (see Fig. ~\ref{plot_topo}).
%%%%%%%%%%% Ici
 { Additionally, adopting a higher order description (lower curve in Fig. ~\ref{plot_topo})
 may introduce false features
 like local peaks. The locally linear model (upper curve in Fig. ~\ref{plot_topo}) is the
 simplest description consistent with the general properties of $\frac{d \tilde f_0}{d \theta}$;
it has the smallest possible number of minima and maxima. Thus adopting the locally 
linear model guarantees that no un-necessary features are introduced in the model. 
On the other hand, in reality this description may be too simple. The validity of this model
can only be tested by a direct application to the data, and an evaluation of the quality of the fit.
 A higher order description should be adopted only if the locally linear model does not fit the data properly.}
{ As a consequence, the locally linear model will be adopted in the continuation of this work.}
Once the general shape of the solution
is known, everything is now just a matter of improving the numerical accuracy.
The process of building the solution will proceed in several steps. First the local
solution will be refined numerically, then some interpolation of the solution in
dark areas will be computed. A Fourier expansion will be fitted
to this piecewise polynomial solution. And finally the adjustment of
the Fourier expansion to the data will be optimized
 using a non linear minimization procedure. Note that
since only the square of $\frac{d \tilde f_0}{d \theta}$ enters
Eq. (~\ref{dr_eq}), the sign of $\frac{d \tilde f_0}{d \theta}$ is not known, as
a consequence two solutions with different sign will have to be
explored. Hopefully, when the source is not circular this degeneracy of the sign can
be broken.
\subsection{Refinement of the local solution.}
\begin{figure}
\centering{\epsfig{figure=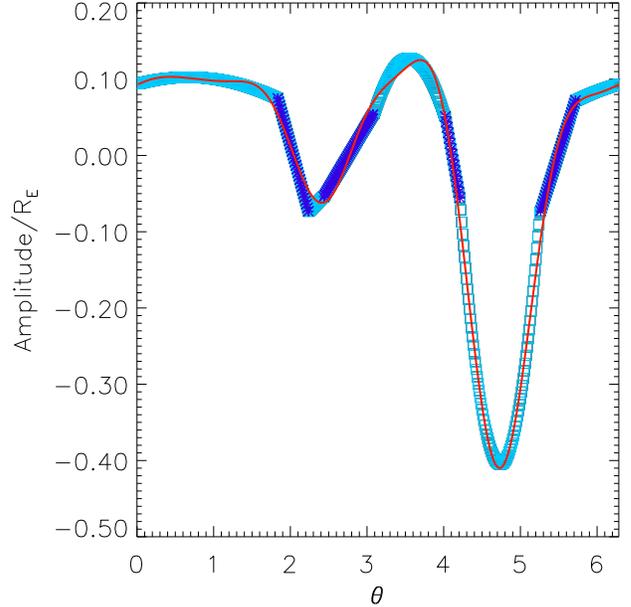,width=9cm}}
\caption{An estimate of the $\frac{d \tilde \tilde f_0}{d \theta}$ field. The 
  blue line segments represent the local linear guess at the image
  location, while the light blue curves are the second order
  interpolations between images. The red line is an adjustment of a 6th order Fourier
  series to the blue curve.}
\label{plot_guess}
\end{figure}

The study of the general properties of $\frac{d \tilde f_0}{d \theta}$ indicates
that the simplest model of this field is locally linear in the neighborhood of the
images. This local linear model will be refined in this section.
The same local linear approximation will be used for the field $\tilde f_1$. 
As illustrated in the former section, the estimation will start from the circular source guess.
For a circular source and a linear field model, 
the field $\frac{d \tilde f_0}{d \theta}$ is zero 
at the center of the image. This property will be assumed in the
refinement of the local solution.
The slope of the linear model for $\frac{d \tilde f_0}{d \theta}$ 
is estimated by solving
Eq. (~\ref{dr_eq}). Note that solving Eq. (~\ref{dr_eq}) requires an 
estimation of the source radius $R_0$. Since the image maximum half
radial width is $R_0$, an estimation of $R_0$ can be obtained by
averaging the maximum radial width of the 4 images. By adopting this linear
local model of $\frac{d \tilde f_0}{d \theta}$, taking the mean
radial position as an estimation of $\tilde f_1$ linear parameters,
and using the perturbative 
lens equation (Eq. ~\ref{final_pert_eq}), it is possible to fold the 4 images 
to the source plane. Once folded to the source plane, the 4 images must be similar and their
comparison is a direct test of the quality of the lens re-construction. With the initial guess for the
field a,
correlation between the 4 folded images is effectively observed: the mean
cross-correlation between the images is 0.6. This cross
correlation can be improved quite significantly by non-linear
optimization. The cross-correlation between images in the source
plane is maximized using the Nelder \& Mead (1965) simplex method. 
The parameter space of the maximization is the 4 local slopes of the
$\frac{d \tilde f_0}{d \theta}$ field, and also the 4 local slopes of the
field $\tilde f_1$. The simplex maximization brings the mean
cross-correlation between images to 0.8 which is a good indication
that the linear approximation is already a good model. This maximization
is also the occasion to build a first realistic model of the source, and to
go beyond the circular source approximation. The result for
the field $\frac{d \tilde f_0}{d \theta}$ is presented in
Fig. (~\ref{plot_guess}). The local linear approximation is the dark
blue line segments in Fig. (~\ref{plot_guess}).
\subsection{Interpolating the local linear solution in dark areas.}
Since there are no constraints on the field $\tilde f_1$ in dark areas, the
fit of this field in the full angular range is straightforward.
Due to the constraint that no images are formed in dark areas, the
situation is more complicated for the field $\frac{d \tilde f_0}{d \theta}$.
The local linear solution obtained in the former section needs to
to be completed with second order polynomials to fill the gaps
between the images.
The condition that no images are formed in dark areas (Diego {\it et
  al.} 2005) is equivalent to the condition that the
functional $|\frac{d \tilde f_0}{d \theta}|$
is larger in dark areas than in areas of image formation.
Another constraint is that the sum
of the field $\frac{d \tilde f_0}{d \theta}$ must be zero. And finally
the field must be continuous and as smooth as possible. The first
constraint is a constraint on the sign of the curvature of the local
polynomial, or equivalently on the sign of the second order
coefficient of the
polynomial. The second constraint reduces the number of unknowns in the
polynomial coefficients. And the last constraint 
settles the value of the remaining coefficients. Note that the
constraints on zero and first order polynomial coefficients can be
solved using linear methods, which reduce the number of unknown and
leaves us with variables depending only on the second order coefficients.
The remaining equations depending on the second order coefficients have
to be solved with a constraint on the sign of the coefficients. A
simple grid method was adopted to solve the relevant equations.
The final solution is presented in Fig. ~\ref{plot_guess} (light
blue curve). 
\subsection{Final fitting of the fields.}
The final fitting of the fields will be conducted by reconstructing
the images of the source and comparing these images with the HST data 
by chi-square estimation. This procedure will take into
account the convolution of the images with the
PSF, which was ignored in the former reconstructions. 
The HST PSF is estimated using the Tiny Tim software (Krist 1995). 
Note that the chi-square value between model and data can be interpreted
as a comparison between the images. In the model, all images come from the same
source, thus the chi-square directly measures the image similarity in
the lens plane. In practice, the fitting procedure will work in the
following way: start from a numerical guess of the field, reconstruct
the source, then reconstruct the images of the source, and finally
estimate the chi-square with respect to the HST data. An essential component in this
procedure is the source reconstruction method, which is illustrated in
the next section.
\subsubsection{Source reconstruction}
At this point the simplest method to estimate the source would be to
use the Warren \& Dye
(2003) method. Unfortunately the source is very complex, and this method
would require using too many basis functions, which in practice
is un-tractable. The following method has
been preferred: images are co-added in the source 
plane using an algorithm that preserves the geometrical properties of
the images elements. The basic idea of this geometrical method is that
pixels in the lens plane are divided in 2 triangles. These triangles
are transported to the source plane and are assumed to remain
triangles.
The triangles in the source plane are projected on a grid and the flux is co-added. The
optimal size of the grid is
estimated by numerical experiments. The noise level in the source
image is estimated by looking at a specific area, and only the
significant pixels in the source grid are kept. The images of the
source in the lens plane are computed using the same geometrical
method, but this time in the opposite direction. The final step
is to compare the images of the source with the data.
The drawback of this method is that the pixels are affected by the convolution
with the PSF, and thus the source reconstruction is not perfect.
However, it is easy to correct this effect by using the following procedure:
first ignore the effect of the PSF on source reconstruction, find a solution, and then
correct the effect using a deconvolution procedure very similar to the
Van-Cittert method (Van Cittert 1931).
{ This de-convolution of the source starts from the solution obtained ignoring the PSF convolution.
The initial guess is convolved with the PSF and a difference image is formed by subtracting
the data. This difference image is taken as a new input to the source re-construction procedure
in order to produce a first correction of the initial guess of the source. This procedure is further
iterated until only noise is present in the difference image. 
In practice this method requires
to control the noise in the difference image. In this work a 3-$\sigma$ level was adopted; tests
were performed with other thresholds, 2 and 4 $\sigma$, with no significant change in the results.
The lower $\sigma$ cuts require more iterations and provide somewhat more noisy source re-construction.
However, since the areas where effective corrections are needed are quite small (these areas correspond
to the bright sharp features in the images), the source reconstruction
is not affected much. It is also
important to note that these PSF corrections have very little effect on the re-construction of the perturbative
fields. By comparing the fit obtained by ignoring the PSF and correcting for the PSF, it was found that the
difference is only a small fraction of the noise in the field re-construction. Thus for this particular lens
this procedure of PSF de-convolution is important only for the re-construction of the source. 
This is due to the fact that for this particular lens the thickness of the arc system is quite 
large with respect to the PSF size, and as a consequence the PSF convolution has only a limited effect.
}
\subsubsection{Fitting of a Fourier model.}
\begin{figure}
\centering{\epsfig{figure=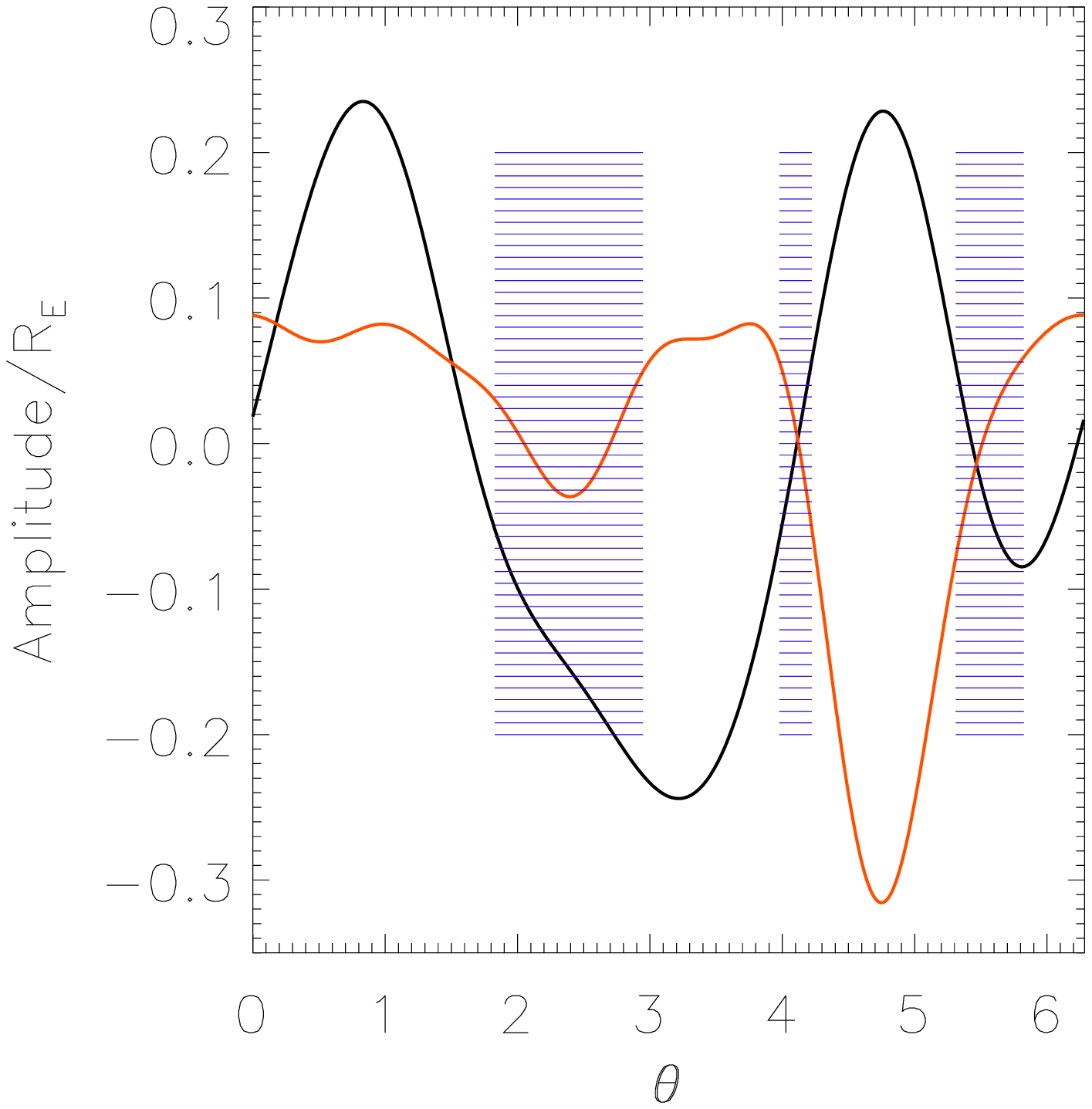,width=9cm}}
\caption{Best solution for the fields: red curve, $\frac{d \tilde \tilde f_0}{d
    \theta}$, and black curve $\tilde \tilde f_1$. The blue dashes indicates areas
    where images are present.}
\label{plot_df0_f1}
\end{figure}

\begin{figure*}
\centering{\epsfig{figure=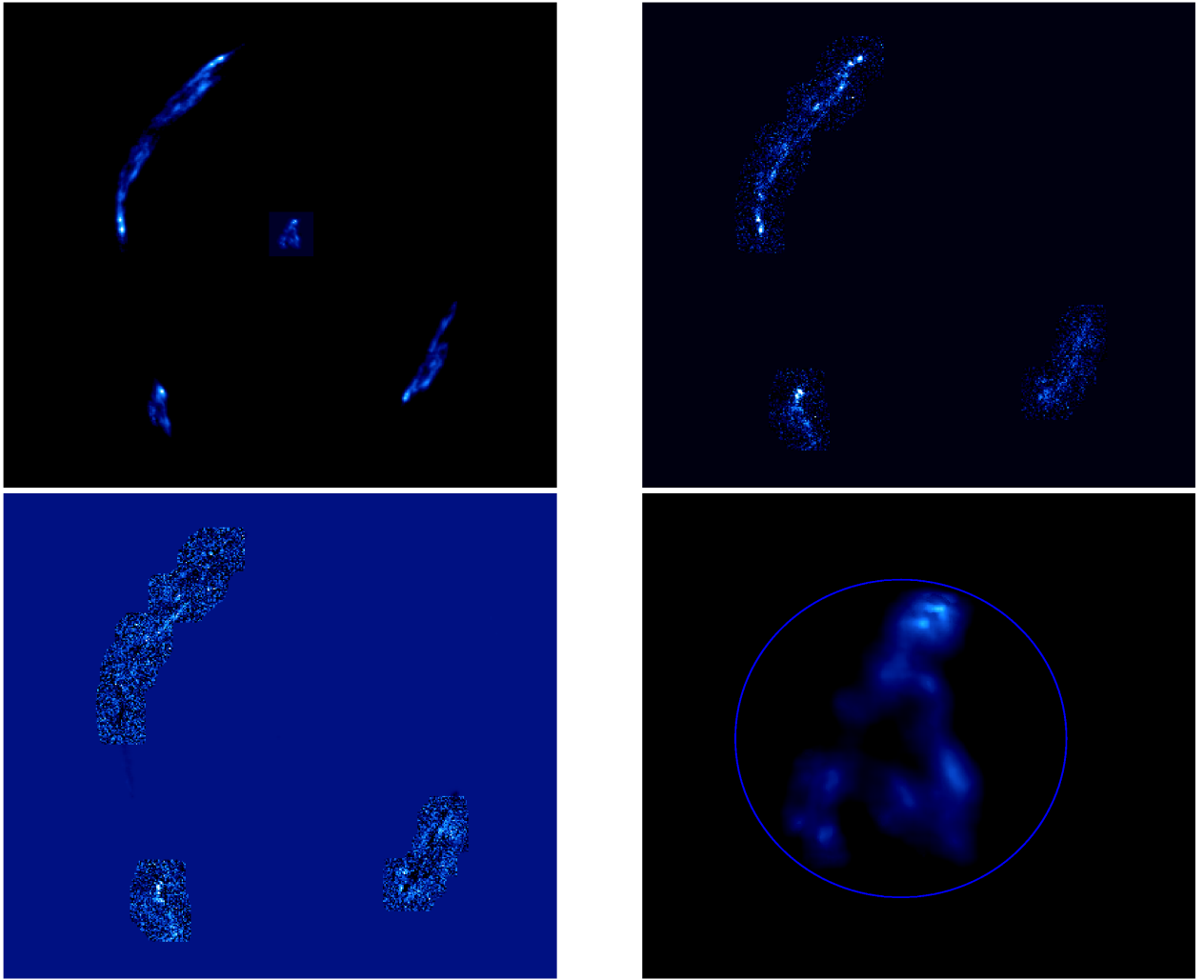,width=18cm}}
\caption{Results from image re-construction: HST data (upper right),
  model (upper left), data and model subtraction (lower left), source
  (lower-right). Note the reproduction of the source details in the
  images of the model. An image of the source
  has been super-imposed on the model (upper left), to illustrate
  the scaling.
}
\label{plot_2}
\end{figure*}

\begin{figure*}[t]
\centering{\epsfig{figure=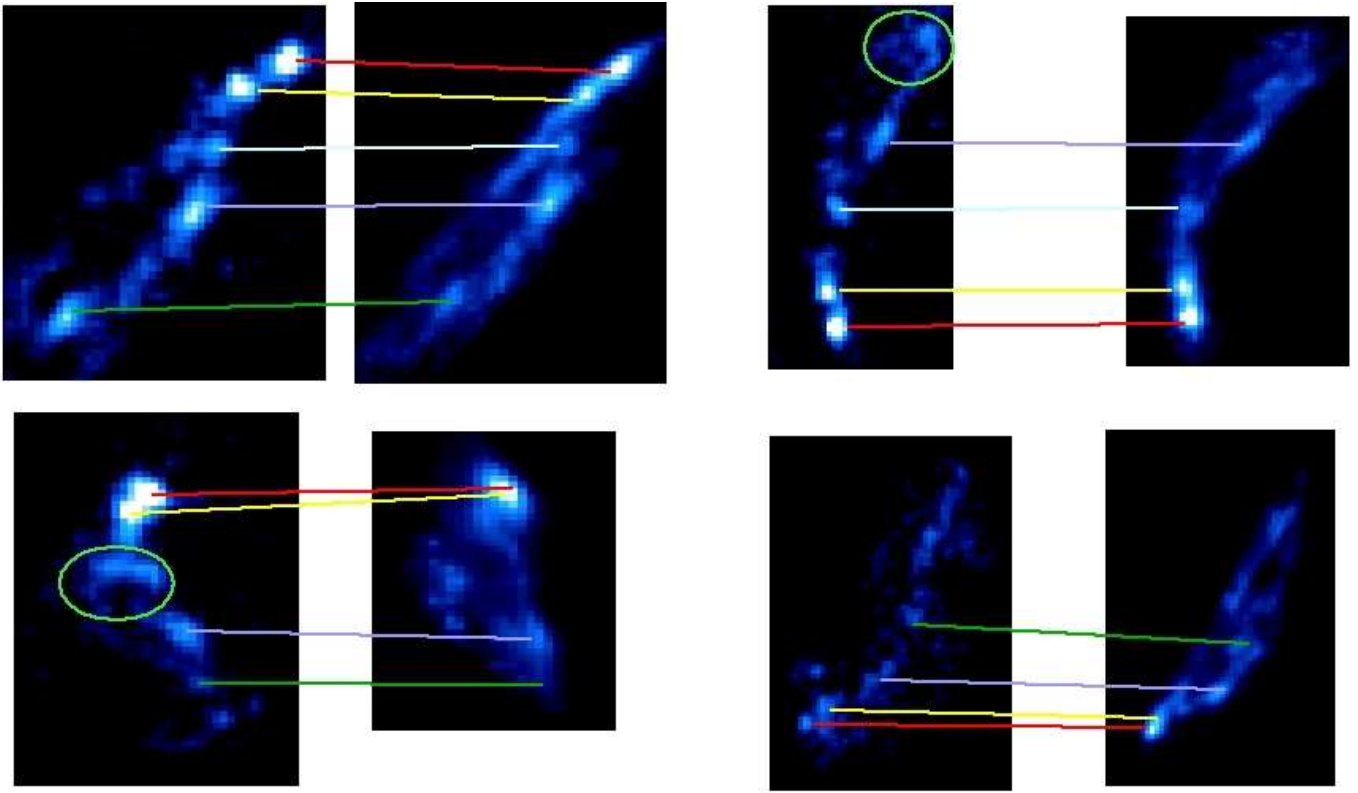,width=18cm}}
\caption{Correspondence between the features in the model (right side) and the HST
data (left side). The color code of the relevant source feature is presented in
Fig. ~\ref{plot_source_details}. The green ellipses indicate areas
where cosmic cleaning was not satisfactory. There are a very large
number of cosmic in the images, and sometimes several of them falls
nearly in the same area, making cosmic cleaning very difficult.}
\label{plot_images_details}
\end{figure*}

\begin{figure}[b]
\centering{\epsfig{figure=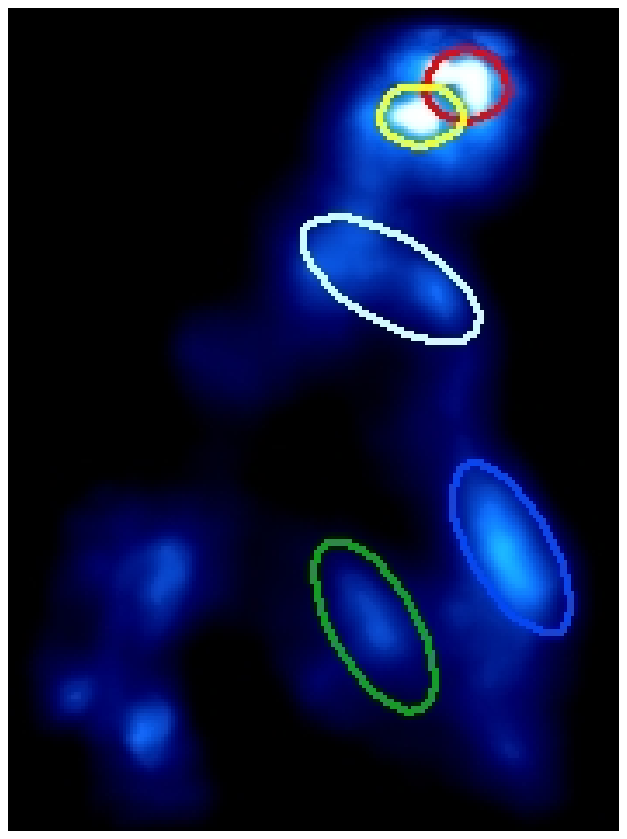,width=5cm}}
\caption{The source features visible in the images
 (see Fig. ~\ref{plot_source_details}). The 3 brightest features
 are outlined by red, yellow, and blue ellipses. These features
 are used to estimate the accuracy of the reconstruction.
 The part of the source at the lower left of the image is
 effectively visible only in one of the images (see Sec. 5 for more
 explanations). }
\label{plot_source_details}
\end{figure}

The practical implementation of the fitting procedure presented above
now will be described. The first step is to choose a starting
estimate. It could be the piecewise polynomial model presented in
Sec. 4.6, but since the Fourier expansion of the fields is directly related to the
multipole expansion of the potential (see Eqs ~\ref{harm_exp}
~\ref{harm_coeffs} and ~\ref{f1f0_harm}), it is more convenient to use
Fourier series. The initial guess is obtained by fitting a 6th order
Fourier series to the piecewise model of $\frac{d \tilde f_0}{d \theta}$, and
for $\tilde f_1$ by fitting a Fourier series to the mean radial position of the images. 
Starting from this initial guess, the source and image reconstruction
procedures are iterated
using the simplex method, in order to minimize the chi-square between
the model and data. This procedure converges to the minimum of the
chi-square in the parameter space. This minimum corresponds to the
optimal Fourier coefficients. Note that as pointed out in Sec. 4.2,
the sign of the initial guess is unknown and 2 solutions with
different signs have been explored. Fortunately in this case the
source is not circular and the 2 solutions have different
  chi-square, which solves the problem of the degeneracy of the
  solution sign. The final solution, the source
reconstruction and the image reconstruction
are presented in  Fig ~\ref{plot_2}, and the field solution is
presented in Fig. (~\ref{plot_df0_f1}).
The fact that the source is complex and has a number of small features
(see Fig. ~\ref{plot_source_details})
can be used to estimate the accuracy of
the modeling. There are 3 bright features in the source that are
visible in all images (red, yellow, blue in Figs
~\ref{plot_images_details} and ~\ref{plot_source_details}).
The position of these features in the HST data and model has been
compared for each image. To obtain the best possible accuracy in the
comparison of the positions, the following procedure was used: a small
image of each feature is extracted in the model frame, and this small
image is cross-correlated with the HST data at various positions
around the feature. The cross-correlation map obtained using this procedure is fitted with a polynomial to 
locate the map maxima, to derive a direct estimate of the
shift between the features. 
%The results are presented in Table (~\ref{table_res}). 
The residuals are somewhat smaller for the first 2
images. In particular a small discrepancy is visible in the bright part of the third
image (lower left image); the distance between the two images of the
bright features in the source are closer in the model. The discrepancy
is of about 2.4 pixels, which is about 1.6 \% of the critical radius.
On average the mismatch between model and observations is
$0.9 \%$ of $R_C$. This estimation is very close to the results obtained
by Peirani et al. (2008) who found that for realistic lens models
the mean errors on the reconstruction of images using the perturbative
approximation was about $1 \%$ of $R_C$. Note that two other features
are visible in the source (see Fig. ~\ref{plot_source_details}); due
to variable levels of amplification, noise, and blurring by other
nearby features, the
images of these features are not always clearly visible in the HST
data. However, when they are visible in the images, their position and
the
correspondence between model and data are indicated in Fig ~\ref{plot_images_details}.
Note that the caustic system associated with the solution is complex and includes a number
of sharp features (see Fig. ~\ref{plot_color}). The complexity of this caustic system is not 
easily visible in the images; this
is due to the fact that the source is large with respect to the small features in the caustic
system.
\subsection{Noise.}
\begin{figure}
\centering{\epsfig{figure=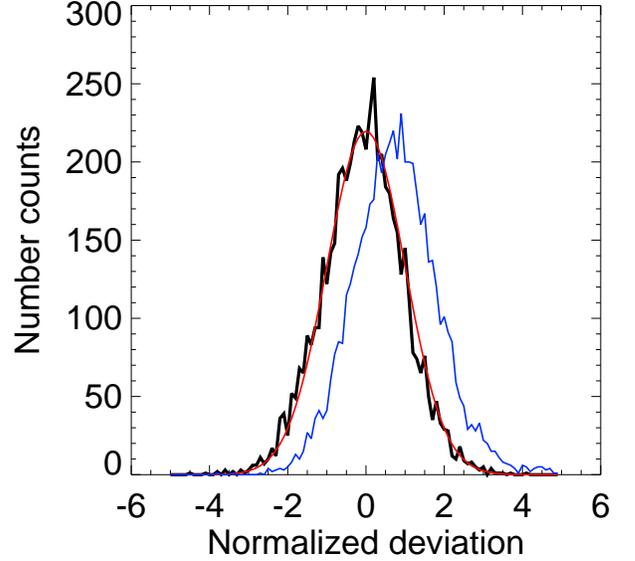,width=8cm}}
\caption{Histogram of the 
  difference image between the model and data. { The black curve is the
  histogram of the pixels in the difference image. For each pixel the difference
  has been weighted by the local Poisson noise expectation. The red curve is the
  theoretical Gaussian expectation for the normalized residuals. The closeness between
  the black and red curves indicates that the result of the fit is very close to optimal. The blue curve 
  is the histogram of normalized residuals of the data without
  subtracting the model. This curve is plotted here for reference; it illustrates the effect of subtracting
  the model from the data.}}
\label{plot_hist}
\end{figure}

 The model and the data
 are compared in a small region around the arcs with total number of pixels $N$.
The Poisson weighted difference $R_i$ between the model at pixel $i$, $M_i$
and the HST data $D_i$ is very close to a Gaussian (see
 Fig. ~\ref{plot_hist}). Considering that the model has $N_P$
 parameters, 25 parameters of the Fourier expansion of the fields,
 and the approximately 450 pixels of the source, the chi-square is,  
 $\chi_{2/dof}=\frac{1}{N-N_P} \sum_i R_i^2 \simeq
 1.17$. 
Changing a little the size of the area, either
 reducing it by moving closer to the center of the arcs, or enlarging
the area does not significantly change the chi-square value.
\subsubsection{Errors on the Fourier coefficients due to noise in the data.}
The errors on the estimation of the Fourier coefficients
are evaluated by local linearization methods near the minima (see, for
 instance Press {et al.} 2007). 
The intensity $M_i$  is a function of the $N_F$ Fourier parameters $p_n$,
$M_i=M_{i}(p_n)$. Assuming that $p_n$ is very close to the solution and
adding a small shift to the parameters $dp_n$, the intensity variation
can be linearized.
\begin{equation}
dM_i=\sum_n \frac{\partial M_i}{\partial p_n} dp_n
\label{lin_err}
\end{equation}
The parameters $dp_n$ can be estimated by a linear least-square fit from
Eq. (~\ref{lin_err}), and thus the relevant errors are also estimated 
by taking the diagonal elements of $A=B^{-1}$, the normal
least-square matrix invert $B$, $B_{nm}= \sum_{i,j} \frac{1}{\sigma_i
  \sigma_j} \left[\frac{\partial M_i}{\partial p_n} \right]
\left[\frac{\partial M_j}{\partial p_m} \right] $. This error estimation
indicates that the errors on the Fourier coefficients due to the Poisson
noise are $\simeq 0.5 \ 10^{-3}$.
\subsubsection{Errors on the reconstruction of the perturbative fields.}
\label{section_pert_err}
In the former section it was shown that the accuracy in the
reconstruction of the images was close to 0.9 \% of $R_C$. This error
is directly related to the errors on the reconstruction of the
perturbative fields. To estimate the amplitude of the errors on the fields
it is particularly useful to consider circular sources.
For a circular source there is a direct relation between the errors on
the fields and the errors on image reconstruction.
Eq. (~\ref{dr_eq}) shows that for circular sources, $\tilde f_1$ is the mean
radial position of the image. As a consequence, the errors on $\tilde f_1$ are of
the same order as the errors on the image position, which were
estimated to be of the order of a percent. The field $\frac{d
  \tilde f_0}{d \theta}$ is related by Eq. (~\ref{dr_eq}) to the radial width
$W$ of the image:
$$
 W^2=R_0^2-\left(\frac{d \tilde f_0}{d \theta} \right)^2 
$$
As a consequence, the error on $W$ and the error on $\frac{d
  \tilde f_0}{d \theta}$ are related:
$$
 \delta W^2 = \delta \left(\frac{d \tilde f_0}{d \theta} \right)^2
$$
Assuming the statistics of the variable $W$ and $\frac{d \tilde f_0}{d
  \theta}$ are the same and differ only by a scale factor,
the above equation shows that this scale factor must be unity.
As a consequence, the errors on $\frac{d f_0}{d \theta}$ are of the same order as the errors
on the image radial width $W$, which are similar to the
  errors on image reconstruction: $\simeq 1\%$.
The corresponding errors on the Fourier coefficients of the field
expansion of order $m$ can be estimated by analyzing the variance of the expansion:
$$
F_i=\sum_{j=1}^m  a_j\cos(j \theta) +  b_j \sin(j \theta)
$$ 
Note that here $F_0=\frac{d \tilde f_0}{d \theta}$ and $F_1=\tilde f_1$.
Taking the square of $F_i$ and averaging over $\theta$
$$
\frac{1}{2 \pi} \int F_i^2 d\theta = \sum_{j=1}^m \frac{a_j^2}{2} + \frac{b_j^2}{2}
$$
Provided it is assumed that the variance of $\langle a_j^2
\rangle=\langle b_j^2 \rangle=\sigma^2$, the
variance of $F_i$ reads
\begin{equation}
 \langle F_i^2 \rangle =  m \sigma^2 \simeq 0.01^2
\label{sigma_fi}
\end{equation}
Note that if the Fourier expansion has no constant term, 
the mean value of $F_i$ is zero ($\int F_i d\theta = 0$).
In practice there is an additional constant term for the field $f_1$, however
considering that the number of Fourier parameters is quite large (12
parameters), the variance due to this additional parameter will be
neglected.  As a consequence for both
fields the errors on the Fourier coefficients is $\sigma \simeq 0.004$. 
This error is about 10 times larger than the error due to the
Poissonian noise in the image, which is clearly negligible. As a
consequence, in the continuation of this work, only the noise
due to the perturbative approximation will be considered.
\subsubsection{Effects of the errors on the potential and density
  estimation.}
Some interesting properties of the lens are directly related to the
perturbative fields as illustrated in Eqs ~\ref{iso_eq} and
~\ref{iso_rho_eq}. The potential iso-contour depends on $f_0$, and
it will be assumed that the variance of $f_0$ is identical to the variance of $\tilde f_0$.
The variance on $\tilde f_0$ can be estimated using the method presented in
Sec. ~\ref{section_pert_err}. The expansion for $\tilde f_0$
 read:
$$
\frac{d \tilde f_0}{d \theta} = \sum_{j=1}^m a_j \cos(j \theta) + b_j \sin(j \theta)
$$
Thus
$$
 \tilde f_0= \sum_{j=1}^m \frac{a_j}{j} \cos(j \theta) + \frac{b_j}{j} \sin(j \theta)
$$
Using Eq (~\ref{sigma_fi}) the associated variance of $\tilde f_0$ is
$$
 \langle \tilde f_0^2 \rangle = \sigma^2 \sum_{j=1}^{m}
 \frac{1}{j^2} = \frac{0.01^2}{m}  \sum_{j=1}^{m}
 \frac{1}{j^2}
$$
As a consequence, the average error on the estimation of $\tilde f_0$ is
$\delta \phi_{iso}\simeq 0.005$.
In the same spirit the errors on the density
iso-contours $\delta \rho_{iso}$ (Eq. ~\ref{iso_rho_eq}) are
$$
 \delta \rho_{iso}^2 =  \frac{0.01^2}{m}  \sum_{j=1}^{\tilde m} \left(j^2+1
 \right)^2
$$
In practice the expansion in the above formula is conducted to order $\tilde m$,
with $1< \tilde m <m$ in order to limit the amplitude of the
noise. Some numerical experiments shows that for $\tilde m>4$ there is no
significant changes to the iso-contour shape, while on the other hand
the noise grows quickly with $\tilde m$. Thus taking $\tilde m=4$, the
resulting error is $\delta \rho_{iso}=0.077$.
%
%%%%%%%%%%%%%% Ici 
%
\section{Prediction of image color.}
\begin{figure*}
\centering{\epsfig{figure=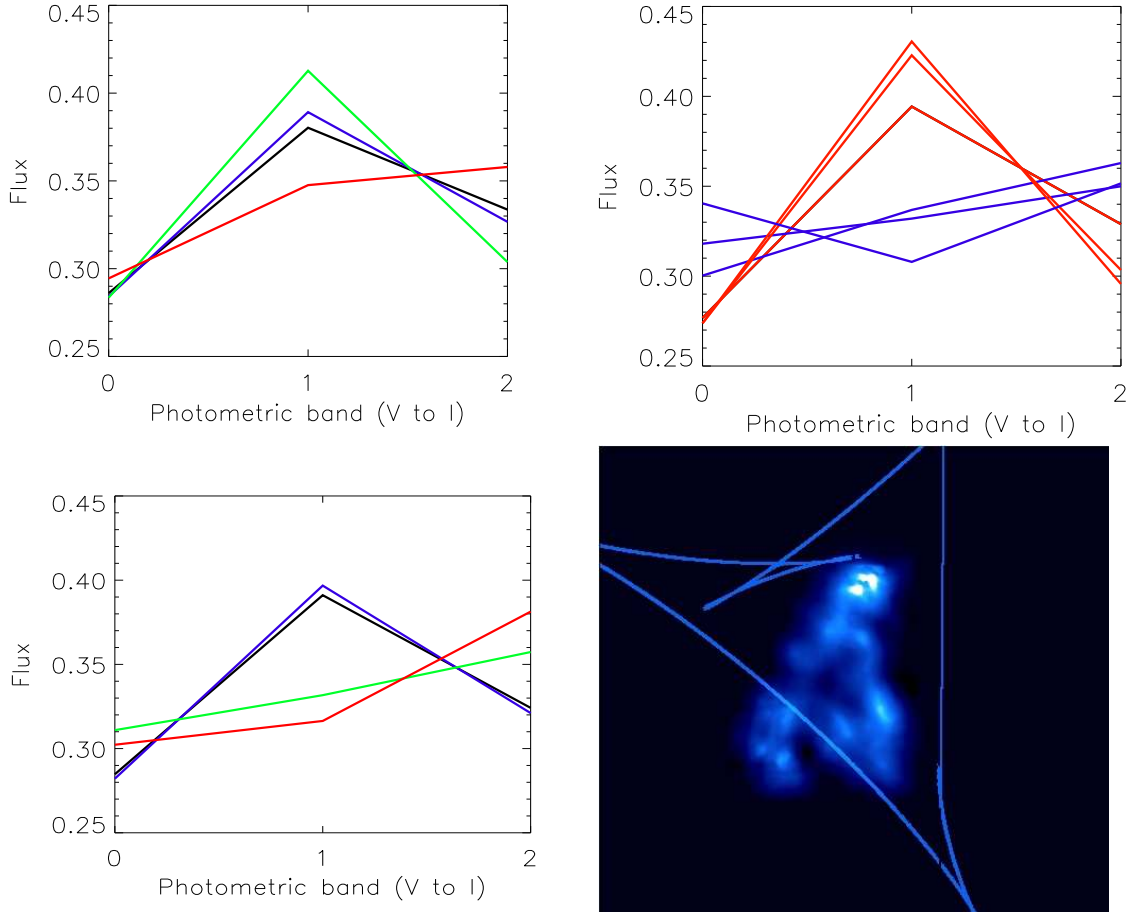,width=16cm}}
\caption{
The 3 color diagrams obtained 
by taking the sequence of raw flux in the 3 HST bands, as well as the
source-caustic system (lower right). 
The upper left presents the color digram of the last image (red),
and of the 3 other images.
The upper right presents the color diagram for the bright
parts of the first 3 images (red curves), while the blue curves
represent the faint parts of the source.
The lower left plot, compares the color diagram of the first 
half of the last image (black curve), to the mean diagram of the 3 other images (blue curve), and
also compares the color diagram of the second
half of the last image (green curve), to the  mean color diagram of
faint source parts in the 3 other images (red curve). The last figure presents the source and the
system of caustics in the source plane.}
\label{plot_color}
\end{figure*}

Figure ~\ref{plot_color} indicates that the last image (lower right position in Fig. ~\ref{plot_1})
does not have the typical "hat-shaped" color diagram 
observed for the 3 other images. 
But this feature is actually a
prediction of the perturbative model. The model predicts that the first half of
the last image comes from
the same source area as the other images. The other half of the
last image corresponds 
to an area of the source that has no counterpart in the large arcs, and only a tiny counterpart
in the lower left image. This situation is due to the crossing of the
caustic line by
a faint part of the source which is visible in Fig. ~\ref{plot_color} in the lower left of
source-caustic diagram. The part of the source situated inside the
caustics corresponds to the first half of the last image, while the
other half of this image corresponds to the faint part of the source
outside the caustics.
As a consequence, the first half of the last
image should have the same color diagram as the other images, and this is
exactly what is observed (see Fig. ~\ref{plot_color}). The remaining
part should have the same color as the faint parts of the source, and
this is also what  Fig. ~\ref{plot_color} illustrates. The color
variations in the source are entirely related to the different colors 
of the bright and faint parts as illustrated in Fig
~\ref{plot_color}. The accurate prediction of the observed image colors
is a confirmation of the model.
\section{Shape of the dark component}
The shape of the potential iso-contour is given by Eq. ~\ref{iso_eq}.
However, in practice, only $\tilde f_0$ is known, and 
 Eq. (~\ref{eq_tilde}) shows that the first
order coefficients of $\tilde f_0$ depend also on the impact
parameters. 
But these first order terms are related to the centering of the potential, not to its shape, for instance
the ellipticity is related to order 2 Fourier terms. 
It is however
possible to obtain some partial information about the centering of the
potential by using Eq. (~\ref{f1f0_harm}) the first order moments
within the unit circle do not depend on the impact parameters (they
cancel out in the calculations). The numerical value of the inner moments
is quite small, of the order of $0.1 R_C$. The outer moments have to
be of the same order, and thus the centering terms are not large.
Note that equations have been normalized by the factor $\kappa_2$
which depends on specific properties of the background.
In practice $\kappa_2$ acts as an unknown scaling factor on $\tilde
f_0$ and thus on the potential iso-contours.
However, an important point is that this scaling degeneracy does not affect the shape
of the potential, and for realistic halos $\kappa_2$ is close to
unity (Peirani {et al.} 2008). 
The potential iso-contours $dr =-\tilde f_0$ are presented in
Fig. (~\ref{plot_pot2}).
%%%%%%%%%%%%%%%%%%%%%% Ici 
\subsection{Mass to light relation}
\begin{figure*}
\centering{\epsfig{figure=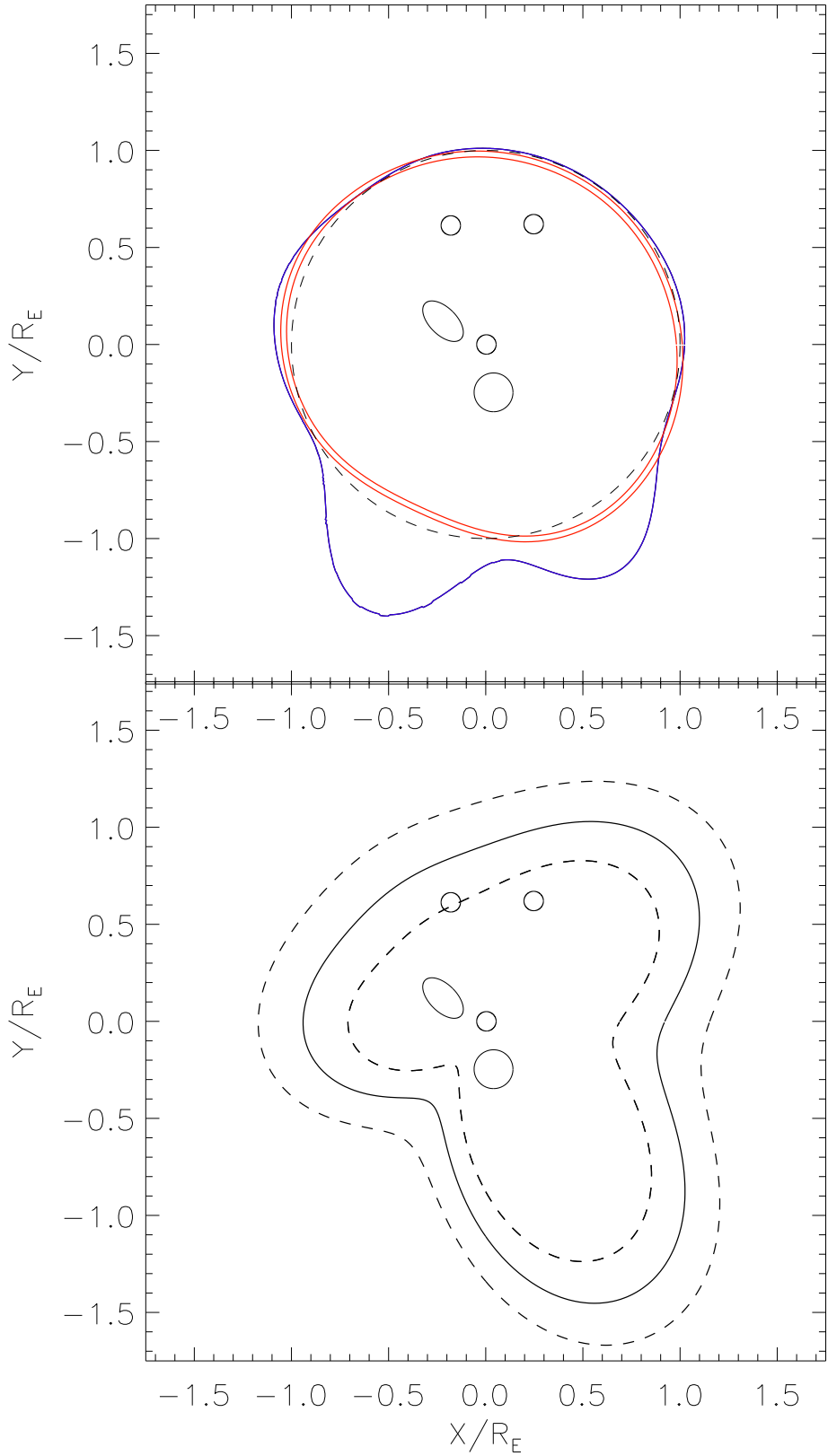,width=12.5cm}}
\caption{The upper plot presents the potential iso-contours, the
  lower plot presents the density iso-contours. In the upper plot, the red
 curves represent the two 3 $\sigma$
  contours of the potential iso-contours. The dashed curve is the unit circle. 
  The blue curve represents the local deviation from circularity of the potential.
The distance between the blue curve and the unit circle
  is proportional to the deviation from circularity.
  The lower plot presents the two 3 $\sigma$ iso-contours of
   the density. In each plot the small elliptical contours represent the galaxies in the lens.
   The small black contours represent the galaxies.}
\label{plot_pot2}
\end{figure*}

The un-perturbed potential iso-contour is a circle with radius $r=1$
when a local perturbation is introduced, some distortions to the circular
iso-contours will be visible in the neighborhood of the perturbator. 
As a consequence, provided that the potential is generated by the
visible matter, the distortions should correlate with the position of
the galaxies. Since the center of the potential is unknown, the
center of the un-perturbed unit circle is also unknown, and will have to be
adjusted. The adjustment of this local circle is performed 
in the neighborhood of points on the potential iso-contour. 
The angular range of the circle adjustment is $\frac{\pi}{2}$.
The residual of this local unit circle adjustment is presented
in Fig. ~\ref{plot_pot2}. The local
 deviations from circularity are not consistent with the positions of
 the galaxies. The opposite is visible; the maximum deviations from
 circularity are observed in the lower part of the figure where
no nearby galaxies are present. The average deviation from circularity
of the potential in this area is $\simeq 0.05$ $R_C$. According to the mean
error on the potential estimated in Sec. 4.6.3, this is a $10 \sigma$ deviation.
This result demonstrates the need for a dark component
 that does not follow light.
 This is an interesting point for theories that try to avoid dark matter
 by modifying gravity. A natural solution is to consider
 that the dark component was formed by merging of cold dark matter
 halos.
\subsection{Inner and outer contribution to the potential.}
Note also that Eq. ~\ref{f1f0_harm}
relates the Fourier coefficients of the fields $\frac{\tilde f_0}{d \theta}$
and $\tilde f_1$ to the inner ($a_n$,$b_n$) and outer terms ($c_n$,$d_n$) of
the multipole expansion. As a consequence, the expansion of $\tilde f_0$
can be separated into inner contributions (which is equivalent to taking
$c_n=d_n=0$ in the original expansion), and outer terms ($a_n=b_n=0$).
A numerical estimation shows that about 75 \% of the total power
spectrum of the field $\tilde f_0$ comes from inner terms, which demonstrates
clearly that the observed effects are dominated by the inner
contributions. Since the deviation from circularity of the potential in
areas with no light are much above the noise ($10 \sigma$), the
result that mass does not follow cannot be affected by the 25 \%
contribution of the outer terms.
%
%
%\begin{figure}
%\centering{\epsfig{figure=12003f10.ps,width=14cm}}
%\caption{The two 3 $\sigma$ envelopes of the potential iso-contours.
%  The dashed curve is the unit circle.
%  The small black contours
%  represents the galaxies. The potential iso-contours are plotted for
%  $\kappa_2=1$. Other values of $\kappa_2$ would corresponds to a
%  re-scaling of the iso-contours.}
%\label{plot_pot1}
%\end{figure}
%
%
\subsection{Density}
To estimate the density of the lens it is necessary to make
some assumptions. Making a specific assumption about the
background allows us to make an estimation of $\kappa_2$ and also
break the mass-sheet degeneracy. For nearly isothermal lens distributions, the density
iso-contours are given by Eq. ~\ref{iso_rho_eq}.
Note that in this equation the terms corresponding to the
impact parameters cancel out
$$
dr_i=\frac{d^2 f_0}{d^2 \theta}+f_1=\frac{d^2 \tilde f_0}{d^2 \theta}+ \tilde f_1
$$
%The coefficient $\kappa_2$ disappear also in the nearly
%isothermal expansion ($\kappa_2=1+\epsilon \delta \kappa_2 $).
To optimize the signal to noise ratio, the iso-contours presented in Fig. ~\ref{plot_pot2} are reconstructed
by using an order 4 Fourier expansion (see Sec. 4.6.3 for details). Note that a method to
reconstruct the
iso-contours of a nearly isothermal lens was also studied by Evans \& Witt
(2003). However, their description of the potential is less general
since it contains only one angular functional, it is suitable
only for potentials with constant isophotes.
%
%
%\begin{figure}
%\centering{\epsfig{figure=density2.ps,width=14cm}}
%\caption{Estimation of the projected density iso-contour near the
%  critical radius. The dashed lines are the 3 $\sigma$ envelopes.}
%\label{plot_dens}
%\end{figure}
%
%
\section{Discussion}
This paper describes the structure of the dark matter envelope
of a small group of galaxies. It is shown that the distribution
of the dark mass does not follow light (Fig. ~\ref{plot_pot2}).
This is direct evidence for an independent dark component 
at the scale of galaxies. Such new results about a system
with a mass much lower than a typical cluster (for instance the Bullet
Cluster) are essential. 
First because there has always been a problem
with missing baryonic mass in clusters (for instance X ray gas);
second because structures
like the Bullet Cluster contains many galaxies, and their distances
are not known with good enough accuracy to avoid some degeneracy
 (Clowe { et al.} 2006). 
The galaxies in this small group 
have very much the same colors, and as a consequence are probably
very nearly at the same distance. Interpretating these results
is not ambiguous and we are probably observing a merging event
between two cold dark matter halos. Since the dark matter halos are
much more extended than the bright cores of the galaxies, the
interaction between the dark halos is already quite strong, while
at the same time the stellar distribution is much less affected. 
However, by looking carefully at the outer part of the luminous halo's
of the galaxies, some indications of interaction are visible. In
particular, the lack of asymmetry in the outer contours of the light
distribution indicates that the small galaxy at the center of the
group has probably been stripped of its outer parts. This issue will
be analyzed in more detail in a forthcoming paper.
The analysis of this particular lens makes it clear that the combination of the perturbative method and good
HST data open interesting new possibilities. Using this approach
it should be possible to probe the structure of the dark component at
the mass scale of galaxies for a large number of systems. Statistical
results about the geometry and structure of the dark halos should be
derived, and will probably offer the possibility to evaluate the
amount of substructure in the halos (Alard 2008). It is clear also
that the mass-luminosity relation could be analyzed for a number of
other systems, and that the discovery of systems similar to this one
 would constrain even more a modification of 
gravity.  
%
%
%%%%%%%%%%%%%%%%%%%%%%%%%%%%%%%%%%%%%%%%%%%%%%%%%%%%%%%%%%%%%%%%%%%%%%%%%%%%%%%%%%%%%%%%%%%%
%
% Appendice section
%
%%%%%%%%%%%%%%%%%%%%%%%%%%%%%%%%%%%%%%%%%%%%%%%%%%%%%%%%%%%%%%%%%%%%%%%%%%%%%%%%%%%%%%%%%%%%
%
%
\begin{appendix}
\section{Physical meaning of the perturbative approximation.}
\begin{figure}
\centering{\epsfig{figure=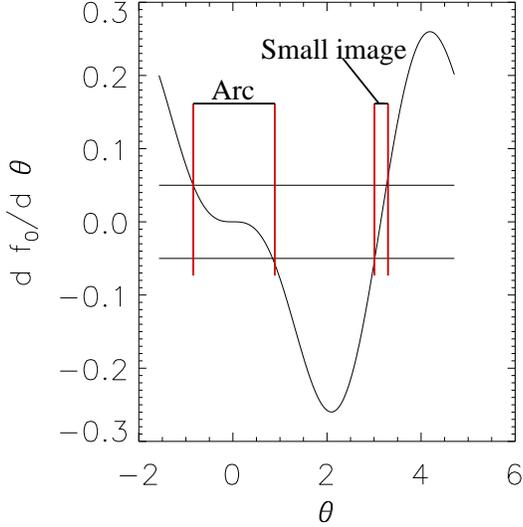,width=10cm,bbllx=0pt,bblly=0pt,bburx=750pt,bbury=750pt}}
\caption{Plot of the function $\frac{d \tilde f_0}{d \theta}$ ($x_0=2 \eta$ and $y_0=0$ in Eq. ~\ref{ex_eq}). Images are
formed when $\frac{d \tilde f_0}{d \theta}<R_0$. The nature of the image is related to the
behavior of  $|\frac{d \tilde f_0}{d \theta}|$ near its minima, the arc corresponds to a large minima, with cancellation
of the first and second order derivatives of $\frac{d \tilde f_0}{d \theta}$. For the small image, the behavior of  $\frac{d \tilde f_0}{d \theta}$ near zero
is linear, and corresponds to a much smaller minima of  $|\frac{d \tilde f_0}{d \theta}|$.}
\label{plot_expl2}
\end{figure}

For circular sources, Eq (~\ref{dr_eq}) shows that $\tilde f_1$ is the mean radial position of the image.
The width of the image is $\sqrt{R_0^2-\left( \frac{d \tilde f_0}{d \theta} \right)^2}$.
The condition for image formation is $|\frac{d \tilde f_0}{d
  \theta}|<R_0$, 
thus images are formed near the minimas
of $|\frac{d \tilde f_0}{d \theta}|$. The minimas of $|\frac{d \tilde f_0}{d \theta}|$ are of two different
types: linear behavior of $\frac{d \tilde f_0}{d \theta}$ near
the minima (small images), or a higher order polynomial
behavior: caustics. To illustrate these properties, let's take the simple case of an elliptical potential,
with ellipticity parameter $\eta$:
$$
\phi=F \left (\sqrt{(1-\eta) x^2+(1+\eta) y^2} \right)
$$
Let's now develop this potential to first order in $\eta$:
\begin{equation}
\phi \simeq F(r)-\frac{\eta}{2} F^{\prime}(r) r cos 2 \theta
\label{ell_pot}
\end{equation}
As a consequence, $\phi_0(r)=F(r)$, $\psi(r,\theta)=\frac{\eta}{2} F^{\prime}(r) r cos 2 \theta$ (see Eq. ~\ref{phi_pert}) and
 since the background potential must be critical at $r=1$, $F^{\prime}(1)=1$. Using Eqs (~\ref{fc_defs}) and (~\ref{eq_tilde}) the perturbative
fields reads:
\begin{equation}
 \begin{cases}
  \tilde f_1=-\frac{\eta}{2} \left(1+F^{\prime \prime}(1)\right) \cos 2 \theta+x_0  \cos \theta +y_0 \sin \theta\\
    \frac{d \tilde f_0}{d \theta}=\eta \sin 2 \theta-x_0  \sin \theta +y_0 \cos \theta
 \end{cases}
 \label{ex_eq}
\end{equation}
Placing the source on the X-axis ($y_0=0$), at an impact parameter $x_0=2 \eta$ Eq. (~\ref{ex_eq}) 
one obtains at $\theta=0$: $\frac{d^n f_0}{d\theta^n}=0$, $n=1..3$,
which corresponds to a broad minima of 
$|\frac{d \tilde f_0}{d \theta}|$ near $\theta=0$  (see Fig. ~\ref{plot_expl2}).
{ This minimum corresponds to the formation of an arc.}
This is a case of a cusp caustic, and
$\frac{d \tilde f_0}{d \theta}$ behaves like an order 3 polynomial near the singularity.
{ The counter-image in the opposite direction is much smaller }
and corresponds to a narrow minimum of $|\frac{d \tilde f_0}{d \theta}|$.  For
this small image $\frac{d^n f_0}{d\theta^n}=0$ for $n=1$
only, and the behavior of this field is linear near the image.
%
%\begin{figure}
%\centering{\epsfig{figure=expl1.ps,width=10cm,bbllx=0pt,bblly=100pt,bburx=0pt,bbury=100pt}}
%\caption{Illustration of a cusp caustic arc ($x_0=2 \eta$ and $y_0=0$ in Eq. ~\ref{ex_eq}), and the associated small
%  image in the opposite direction. The source is the black dot between
%  the 2 images. The 2 images of the source are obtained by ray-tracing,
%  and the the red contours are the images of the outer contour of the
%  source obtained using the perturbative approximation.
%In this example, $F^{\prime
%  \prime}(1)=0$ was adopted.
%The formation of these 2 images is easy to understand in the framework of perturbative theory (see Fig.
%~\ref{plot_expl2}).}
%\label{plot_expl1}
%\end{figure}
%
%
\label{appendice_1}
\end{appendix}
%
%%%%%%%%%%%%%%%%%%%%%%%%%%%%%%%%%%%%%%%%%%%%%%%%%%%%%%%%%%%%%%%%%%%%%%%%%%%%%%%%%%%%%%%%%%%%%%%%%%%%%%%%%%
%%%%%%%%%%%%%%%%%%%%%%%%%%%%%%%%%%%%%%%%%%%%%%%%%%%%%%%%%%%%%%%%%%%%%%%%%%%%%%%%%%%%%%%%%%%%%%%%%%%%%%%%%%
%
\begin{appendix}
\section{Relation to multipole expansion}
 The expansion of the perturbative potential { at a given radial position $r$ } $\psi$ reads (Kochanek 1991):
\begin{equation}
 \begin{aligned}
 \psi = -\sum_n  \frac{a_n(r)}{r^n} \cos n \theta + \frac{b_n(r)}{r^n} \sin n \theta + \\
   c_n(r) \ r^n \cos n \theta + d_n(r) \ r^n \sin n \theta  \\
 \end{aligned}
 \label{harm_exp}
\end{equation}
The perturbative theory requires the evaluation of the potential gradient on the critical circle. 
 Since the coordinate system has been re-normalized so that the critical radius is situated at $r=1$,
all the evaluation of quantities related to the potential will be performed at $r=1$.
The coefficients $(a_n,b_n,c_n,d_n)$ are related to the projected density of the lens $\rho$ by the following formula:
\begin{equation} 
\begin{cases}
a_n = \frac{1}{2 \pi n} \int_0^{2 \pi} \int_0^{r=1} \rho(u,v) \cos n v  \ u^{n+1} \ du \ dv \\
b_n = \frac{1}{2 \pi n} \int_0^{2 \pi}  \int_0^{r=1} \rho(u,v) \sin n v  \ u^{n+1} \ du \ dv \\
c_n = \frac{1}{2 \pi n} \int_0^{2 \pi} \int_{r=1}^\infty \rho(u,v) \cos n v  \ u^{1-n} \  du \ dv \\
d_n = \frac{1}{2 \pi n} \int_0^{2 \pi} \int_{r=1}^\infty \rho(u,v)
\sin n v \ u^{1-n} \ du \ dv \\
\label{harm_coeffs}
\end{cases}
\end{equation}
Using Eqs (~\ref{harm_exp}) and  (~\ref{harm_coeffs}), and noting that:
 $\left(\frac{d \left( a_n+c_n \right)}{d r}\right)_{[r=1]}=\left(\frac{d \left( b_n+d_n \right)}{d r}\right)_{[r=1]}=0$ 
the fields $ \frac{d f_0}{d \theta} = \left(\frac{\partial \psi}{\partial \theta}\right)_{[r=1]}$ and $f_1 = \left(\frac{\partial \psi}{\partial r}\right)_{[r=1]}$ :
 \begin{equation}
  \begin{cases}
  f_1 = \sum_n n \left(a_n-c_n \right) \cos n \theta 
  + n \left(b_n-d_n \right) \sin n \theta\\
  \frac{d f_0}{d \theta} = \sum_n -n \left(b_n+d_n \right) \cos n \theta 
  + n \left(a_n+c_n \right) \sin n \theta \\
  \label{f1f0_harm}
 \end{cases}
 \end{equation}
As a consequence, there is a direct relation between the Fourier
expansion of the perturbative fields and the multipole expansion of
the potential at the critical circle. This relation also demonstrates
that the Fourier expansion of the perturbative fields is not
equivalent to the Fourier expansion of the potential by Trotter {et al.}
(2000). The expansion proposed by Trotter requires a number of
additional coefficients with respect to the multipole expansion.
On the other hand, Eqs (~\ref{harm_exp}), (~\ref{harm_coeffs}) and
(~\ref{f1f0_harm})
indicate that the perturbative expansion is directly related to the multipole
expansion without needing any additional coefficients.
It is also clear from Eqs (~\ref{harm_exp}) and (~\ref{f1f0_harm})
that since the Fourier expansion of elliptical potentials is approximately of order 2 (see
Eq ~\ref{ell_pot}), the corresponding Fourier expansion will also be 
of order 2. Higher order terms in the Fourier expansion of the
perturbative fields will thus be related to the skewness,
boxyness, etc. , of the potential. The first order terms are
related to the centering of the potential.
\label{appendix_2}
\end{appendix} 
%
%%%%%%%%%%%%%%%%%%%%%%%%%%%%%%%%%%%%%%%%%%%%%%%%%%%%%%%%%%%%%%%%%%%%%%%%%%%%%%%%%%%%%%%%%%%%%%%%%%%%%%%%
%
\begin{appendix}
\section{Geometrical interpretation}
The perturbative fields are directly related to the geometrical
structure of the potential. The potential iso-contour near the
critical circle is easy to infer from Eq. (~\ref{pot_eq}). The
iso-contour equation reads:
$$
 \phi_0(r)+\epsilon f_0(\theta) +\epsilon f_1(\theta) (r-1)=C
$$
Searching for the iso-contour near the critical circle $r_i=1+\epsilon dr_i$
and developing the former equation to first order in $\epsilon$:
$$
\epsilon \left(dr_i + f_0 \right) = C-\phi_0(1)
$$
When no perturbation is introduced ($f_0=0$), the iso-contour must be
reduced to the critical circle ($dr=0$), as a consequence the right
side of the equation must be zero too, and the former iso-contour
equation reduces to: 
\begin{equation}
 dr_i = -f_0
 \label{iso_eq}
\end{equation}
The potential iso-contour depend only on $f_0$, while the iso-contour
variation depends also on $f_1$. Considering a contour with
radial position $r_i$,
defined by $\phi(r_i,\theta)=C$, the equation of a nearby contour 
is $\phi(r_i+\delta r_i,\theta)=C+dC$, or equivalently $\frac{d \phi}{d r}(r_i,\theta)
\delta r_i = dC$. There is no iso-contour variation provided that $\delta r_i$ is
constant, which is equivalent to $\frac{d \phi}{d r}(r_i,\theta)$
being constant. To first order in $\epsilon$:
\begin{equation}
 \frac{d \phi}{d r}(r_i,\theta) \simeq 1+f_1-\frac{d^2 \phi_0}{d^2 r}(1) f_0
\label{iso_var}
\end{equation}
Considering that Eq (~\ref{iso_var}) depends only on $f_1$ for purely
isothermal backgrounds, and that CDM halos are quite close to
isothermal in general, the iso-contour variation is dominated by $f_1$.
\label{appendix_3}
\end{appendix}
%
%%%%%%%%%%%%%%%%%%%%%%%%%%%%%%%%%%%%%%%%%%%%%%%%%%%%%%%%%%%%%%%%%%%%%%%%%%%%%%%%%%%%%%%%%%%%%%%%%%%%%%%%
%
\begin{appendix}
\section{Relation to the geometry of the density}
The two perturbative fields are related to the potential by the following
 expansion (Eq. ~\ref{phi_pert}):

 \begin{equation}
 \phi=\phi_0(r)+\epsilon f_0(\theta)+\epsilon f_1(\theta) (r-1) 
 \label{pot_eq}
\end{equation}
Note that since $r=1+\epsilon dr$ the expansion presented in
Eq. (~\ref{pot_eq}) is of order 2 in $\epsilon$. This expansion
reduces to order 1 when introduced in the lens equation. 
By inserting Eq (~\ref{pot_eq}) in the Poisson equation, taking
$r=1+\epsilon dr$, and developing to first order in $\epsilon$, the
density reads
$$
\rho \simeq (1+2 C_2) +\left( (2 C_2+6 C_3-1) dr + \frac{d^2 f_0}{d^2
  \theta}+f_1+2 f_2 \right) \epsilon
$$
The iso-contour near the critical circle is defined by the equation
$\rho = C$. In the absence of a perturbation ($f_i=0$) the
iso-contour must reduce to the critical circle ($dr=0$). As a
consequence the equation of the iso-contour is:
$$
 (1-2 C_2-6 C_3) dr_i = \frac{d^2 f_0}{d^2 \theta}+f_1+2 f_2
$$
For a nearly isothermal distribution, the higher order derivatives
($n>1$) are close to zero. Assuming that these higher order
derivatives are of order $\epsilon$ ($C_2 \simeq C_3 \simeq \epsilon$
and $f_2 \simeq \epsilon F(\theta)$), the former equation reduce to:
\begin{equation}
 dr_i = \frac{d^2 f_0}{d^2 \theta}+f_1
\label{iso_rho_eq}
\end{equation}
\label{appendix_4}
\end{appendix}
\begin{acknowledgements}
This work is based on HST data, credited to STScI and
prepared for NASA under Contract NAS 5-26555.
I would like to
thank Fran\c cois Sevre, Leslie Sage, J.P. Beaulieu, S. Mao, R. Blandford, and P. Alard for helping with this paper,
and the referee for helping to improve the presentation of this paper.
\end{acknowledgements}

\end{document}